\documentstyle[preprint,aps,epsf,floats,tighten]{revtex} 

\begin{document}

\preprint{\tighten \vbox{\hbox{hep-ph/0011199}
              \hbox{}\hbox{}\hbox{}\hbox{}\hbox{}\hbox{}\hbox{} }}

\title{New Unitarity Triangles of Quark Mixing and CP Violation}

\author{Changhao Jin}

\address{School of Physics, University of Melbourne\\
Victoria 3010, Australia}

\maketitle

{\tighten
\begin{abstract}%
Unitarity of the Cabibbo-Kobayashi-Maskawa matrix gives rise to nine new
unitarity triangles in the standard model.
\end{abstract}
}

\newpage

Huge samples of $B$ mesons produced at the $B$ factories and recent 
developments in the experiments with kaons and charmed mesons boost the 
studies of CP violation and quark mixing. According to the six-quark scheme of 
Kobayashi and Maskawa \cite{km}, CP symmetry
is violated because one complex phase remains in the 
coupling constants of the charged current weak interactions after taking into
account unitarity and rephasing quark fields. In the standard model CP 
violation and quark mixing are intimately related and described by the 
Cabibbo-Kobayashi-Maskawa (CKM) matrix \cite{km,cabi}
\begin{equation}
V= \left( \begin{array}{ccc}
V_{ud} & V_{us} & V_{ub} \\
V_{cd} & V_{cs} & V_{cb} \\
V_{td} & V_{ts} & V_{tb}   \end{array}
\right).
\label{eq:ckm}
\end{equation}
The CKM matrix $V$ originating from the diagonalization of quark mass 
matrices must be unitary: $VV^\dagger=V^\dagger V=1$. 
Unitarity of the CKM matrix
is at the basis of the GIM mechanism \cite{gim}.
Unitarity strongly constrains and correlates the magnitudes and phases of the 
nine elements 
of the CKM matrix, although their values are not predicted in detail. 
Consequently, CP-conserving observables and CP-violating observables are
connected with each other. Confrontation of unitarity relations of the 
CKM matrix with
experimental measurements is an important test of the standard model.
A deviation from expectations would signal new physics.

Unitarity of the CKM matrix is conveniently displayed in the well-known
``unitarity triangle'' \cite{triangle}. There are six unitarity triangles resulting
from the orthogonality between each pair of rows, or each pair of columns, 
in the $3\times 3$ CKM matrix. 

The orthogonality condition between two different rows reads
\begin{equation}
V_{\alpha d}^\ast V_{\beta d}+V_{\alpha s}^\ast V_{\beta s}+
V_{\alpha b}^\ast V_{\beta b}= 0. 
\label{eq:row}
\end{equation}
Here and hereafter, Greek subscripts ($\alpha, \beta, \gamma$) run over the 
up-type quarks $u, c$, and $t$, while Latin ones ($i, j, k$) run over the 
down-type quarks $d, s$, and $b$.
$\alpha\beta\gamma$ is some cyclic permutation of $uct$, and $ijk$ of $dsb$.
The relation (\ref{eq:row}) can be geometrically represented as a triangle in 
the complex plane, formed by the three sides $|V_{\alpha d}V_{\beta d}|$, 
$|V_{\alpha s}V_{\beta s}|$, and $|V_{\alpha b}V_{\beta b}|$. A physical 
observable must be independent of phase convention of quark fields. Let
\begin{equation}
\omega_{\gamma k}\equiv {\rm arg}(-V_{\alpha i}^\ast V_{\beta j}^\ast
V_{\alpha j}V_{\beta i})
\label{eq:def}
\end{equation}
be the phase of the product of the matrix elements 
$V_{\alpha i}^\ast V_{\beta j}^\ast V_{\alpha j}V_{\beta i}$. Since
$V_{\alpha i}^\ast V_{\beta j}^\ast V_{\alpha j}V_{\beta i}$ is invariant 
under rephasing of quark fields,  $\omega_{\gamma k}$ is an observable phase. 
Then for the three interior angles of the unitarity triangle, 
the angle between the sides 
$|V_{\alpha s}V_{\beta s}|$ and $|V_{\alpha b}V_{\beta b}|$ is 
$\omega_{\gamma d}$; the angle between the sides 
$|V_{\alpha b}V_{\beta b}|$ and $|V_{\alpha d}V_{\beta d}|$ is 
$\omega_{\gamma s}$; the angle between the sides 
$|V_{\alpha d}V_{\beta d}|$ and $|V_{\alpha s}V_{\beta s}|$ is 
$\omega_{\gamma b}$. By definition, 
\begin{equation}
\omega_{\gamma d}+\omega_{\gamma s}+\omega_{\gamma b}=\pi,
\label{eq:sum1} 
\end{equation}
since 
\begin{equation}
(-V_{\alpha s}^\ast V_{\beta b}^\ast V_{\alpha b}V_{\beta s})
(-V_{\alpha b}^\ast V_{\beta d}^\ast V_{\alpha d}V_{\beta b})
(-V_{\alpha d}^\ast V_{\beta s}^\ast V_{\alpha s}V_{\beta d})
=-|V_{\alpha d}V_{\alpha s}V_{\alpha b}V_{\beta d}V_{\beta s}V_{\beta b}|^2.
\label{eq:product}
\end{equation}
The relation (\ref{eq:row}) leads to three unitarity
triangles. We label them by their $\alpha\beta$ indices (with $\alpha\beta$ 
being $uc$, $ct$, and $tu$, respectively).

Similarly, the orthogonality condition between two different columns reads
\begin{equation}
V_{ui}^\ast V_{uj}+V_{ci}^\ast V_{cj}+V_{ti}^\ast V_{tj}= 0.
\label{eq:column}
\end{equation}
This relation can also be geometrically represented as a triangle in the 
complex plane, with three sides $|V_{ui}V_{uj}|$, $|V_{ci}V_{cj}|$, and  
$|V_{ti}V_{tj}|$. For the three interior angles,
the angle between the sides $|V_{ci}V_{cj}|$ and $|V_{ti}V_{tj}|$ is 
$\omega_{uk}$; the angle between the sides $|V_{ti}V_{tj}|$ and 
$|V_{ui}V_{uj}|$ is $\omega_{ck}$; the angle between the sides 
$|V_{ui}V_{uj}|$ and $|V_{ci}V_{cj}|$ is 
$\omega_{tk}$. By definition, 
\begin{equation}
\omega_{uk}+\omega_{ck}+\omega_{tk}=\pi.
\label{eq:sum2}
\end{equation} 
The relation (\ref{eq:column}) also leads to three unitarity triangles. We
label these three triangles by their $ij$ indices (with 
$ij$ being $ds$, $sb$, and $bd$, respectively). Among them is the $bd$
unitarity triangle which is particularly useful for the study of $B$ 
physics, with its sides determined by semileptonic $B$ decays and 
$B^0-\bar{B^0}$ mixing and its angles $\omega_{us}=
{\rm arg}(-V_{cb}^\ast V_{td}^\ast V_{cd}V_{tb})=\beta$, $\omega_{cs}=
{\rm arg}(-V_{tb}^\ast V_{ud}^\ast V_{td}V_{ub})=\alpha$, and 
$\omega_{ts}={\rm arg}(-V_{ub}^\ast V_{cd}^\ast V_{ud}V_{cb})=\gamma$ 
determined by CP-violating asymmetries in $B$ decays. 

Although different in shapes, all the six unitarity triangles have the same 
area being half the Jarlskog invariant \cite{jarlskog} 
$J_{\rm CP}\equiv |{\rm Im}(V_{\alpha i}^\ast V_{\beta j}^\ast 
V_{\alpha j}V_{\beta i})|$.
The Jarlskog invariant $J_{\rm CP}$ is a universal rephasing invariant 
quantity which
is independent of the $\alpha\beta$ and $ij$ indices due to unitarity.  
CP-violating observables are all proportional to $J_{\rm CP}$. 

The above six conventional unitarity triangles picture six orthogonality 
conditions of the CKM matrix. Other different relationships can be obtained
from unitarity of the CKM matrix. Indeed, we find nine new unitarity 
relations following from a combination of orthogonality and 
normalization conditions. These unitarity relations
can also be pictured as triangles with the interesting feature that all the
nine new unitarity triangles also have the same area equal to $J_{\rm CP}/2$. 
Let us demonstrate these unitarity relations. From the orthogonality condition
Eq.~(\ref{eq:row}) between two different rows, we have
\begin{equation}
V_{\alpha i}^\ast V_{\beta i}+V_{\alpha j}^\ast V_{\beta j}=
-V_{\alpha k}^\ast V_{\beta k}.
\label{eq:orth}
\end{equation}
Multiplying Eq.~(\ref{eq:orth}) by its complex conjugation yields
\begin{equation}
2{\rm Re}(V_{\alpha i}V_{\beta j}V_{\alpha j}^\ast V_{\beta i}^\ast)=
-|V_{\alpha i}V_{\beta i}|^2-|V_{\alpha j}V_{\beta j}|^2+
|V_{\alpha k}V_{\beta k}|^2.
\label{eq:conj}
\end{equation}
By applying the normalization condition
\begin{equation}
|V_{\alpha i}|^2+|V_{\alpha j}|^2+|V_{\alpha k}|^2=1 
\label{eq:norm}
\end{equation}
to Eq.~(\ref{eq:conj}), we find
\begin{equation}
2{\rm Re}(V_{\alpha i}V_{\beta j}V_{\alpha j}^\ast V_{\beta i}^\ast)=
2|V_{\alpha i}V_{\beta j}V_{\alpha j}V_{\beta i}|
\cos (\pi-\omega_{\gamma k})=
|V_{\alpha i}V_{\beta j}|^2+|V_{\alpha j}V_{\beta i}|^2-|V_{\gamma k}|^2.
\label{eq:new}
\end{equation}
If we start from the column orthogonality, we arrive at the same relation 
(\ref{eq:new}) by using the normalization condition. The unitarity relation 
(\ref{eq:new}), satisfying the Cosine rule of plane trigonometry, 
defines a triangle with three sides 
$|V_{\alpha i}V_{\beta j}|$, $|V_{\alpha j}V_{\beta i}|$, and $|V_{\gamma k}|$.
Its interior angle between the sides
$|V_{\alpha i}V_{\beta j}|$ and $|V_{\alpha j}V_{\beta i}|$ is 
$\pi-\omega_{\gamma k}$. This is the only angle among its three interior 
angles, 
which is of direct physical interest in the description of CP violation. 
The unitarity relation (\ref{eq:new}) leads to nine unitarity triangles. 
We refer to each of these triangles by the $\gamma k$ indices  
corresponding to its side $|V_{\gamma k}|$. Each of these unitarity triangles
are formed by five magnitudes of the CKM matrix elements, rather than six ones
forming the conventional unitarity triangle. However,
it is easy to show that all these nine new unitarity triangles have the same 
area equal to $J_{\rm CP}/2$ as the conventional ones. A nonzero area implies
CP violation.    

The angles of the unitarity triangles are also correlated. First, the six
conventional unitarity triangles have 9 rather than 18 different angles,
since any triangle representing the row orthogonality (\ref{eq:row})
and any triangle representing the column orthogonality (\ref{eq:column}) 
contain a common angle. This can be understood since one can construct only
nine different phases $\omega_{\gamma k}$ from the rephasing invariants
$V_{\alpha i}^\ast V_{\beta j}^\ast V_{\alpha j}V_{\beta i}$.
On the other hand, there is a one-to-one 
correspondence between these nine different
angles and the nine new unitarity triangles. Second, only 
four out of the total of the nine observable phases $\omega_{\gamma k}$ are 
independent, 
since Eqs.~(\ref{eq:sum1}) and (\ref{eq:sum2}) impose five independent 
constraints on them.
The four independent phases form a complete set of weak phases for 
the description of CP violation, and fully determine the entire CKM 
matrix \cite{alek}. We can choose the four independent phases to be 
$\omega_{us}=\beta$, $\omega_{ts}=\gamma$, 
$\omega_{ud}={\rm arg}(-V_{cs}^\ast V_{tb}^\ast V_{cb}V_{ts})=\chi$, and 
$\omega_{tb}={\rm arg}(-V_{ud}^\ast V_{cs}^\ast V_{us}V_{cd})=\chi^\prime$. 
Using Eqs.~(\ref{eq:sum1}) and (\ref{eq:sum2}), the remaining phases can then 
be expressed in terms of them: 
\begin{eqnarray}
\omega_{ub} &=& \pi-\beta-\chi, \nonumber \\
\omega_{cd} &=& \gamma+\chi^\prime-\chi, \nonumber \\
\omega_{cs} &=& \pi-\beta-\gamma, \nonumber \\
\omega_{cb} &=& \beta+\chi-\chi^\prime, \nonumber \\ 
\omega_{td} &=& \pi-\gamma-\chi^\prime, 
\label{eq:5eqs}
\end{eqnarray}
which are essentially determined by the
two angles $\beta$ and $\gamma$ (in leading order of $\lambda$, except for 
$\omega_{cs}$), since empirically $\beta$ and $\gamma$ are large (i.e., far 
from 0 and $\pi$), while $\chi$ 
is small, of order $\lambda^2$, and $\chi^\prime$ is much smaller, 
of order $\lambda^4$, where $\lambda\equiv |V_{us}|\approx 0.22$.
Note that Eqs.~(\ref{eq:sum1}), (\ref{eq:sum2}) and (\ref{eq:5eqs}) are 
simply a consequence of the definition (\ref{eq:def}) of the phase
$\omega_{\gamma k}$; they have nothing to do with unitarity of the CKM matrix. 
Equations (\ref{eq:sum1}), (\ref{eq:sum2}) and (\ref{eq:5eqs}) still hold, even
if the three family quark mixing matrix is not unitary and the 
unitarity triangle is not closed. Therefore, comparing the relations 
(\ref{eq:sum1}), (\ref{eq:sum2}) and 
(\ref{eq:5eqs}) with experiments does not test unitarity. Deviations from
unitarity will show up in a failure of unitarity relations involving both
angles and magnitudes \cite{silva}.

From Eq.~(\ref{eq:new}) it follows that 
\begin{eqnarray}
\sin \frac{\omega_{\gamma k}}{2} &=& \sqrt{\frac{s_{\gamma k}(s_{\gamma k}-
|V_{\gamma k}|)}{|V_{\alpha i}V_{\beta j}V_{\alpha j}V_{\beta i}|}}\,\,\, , 
\label{eq:new1} \\
\cos \frac{\omega_{\gamma k}}{2} &=& \sqrt{\frac{(s_{\gamma k}-
|V_{\alpha i}V_{\beta j}|)(s_{\gamma k}-|V_{\alpha j}V_{\beta i}|)}
{|V_{\alpha i}V_{\beta j}V_{\alpha j}V_{\beta i}|}}\,\,\, ,
\label{eq:new2}
\end{eqnarray}
where $s_{\gamma k}=(|V_{\alpha i}V_{\beta j}|+|V_{\alpha j}V_{\beta i}|+
|V_{\gamma k}|)/2$ is half the perimeter of the $\gamma k$ triangle. When
the three sides of a unitarity triangle are given, these formulas may be used
in order to find the angle.

The unitarity relations (\ref{eq:new}), (\ref{eq:new1}) and (\ref{eq:new2}) 
can serve as a good test of the
standard model. These tests require precise direct measurements, without
assuming unitarity of the CKM matrix, of the 
sides and angles of the unitarity triangles. By measuring both the sides and  
angles, the unitarity triangles will be overconstrained. Present knowledge of
the magnitudes of the elements in the third row of the CKM matrix 
involving the top quark in Eq.~(\ref{eq:ckm}), but also of $|V_{cs}|$,
$|V_{cd}|$ and $|V_{ub}|$ is still rather imprecise \cite{PDG}. Among the four
independent phase angles $\beta$, $\gamma$, $\chi$, and $\chi^\prime$, only 
direct measurement of $\sin (2\beta)$ has recently been made \cite{beta} 
from the time-dependent CP-violating asymmetry in 
$B^0 \to J/\psi K^0_S$ decays \cite{bisa}. The current experimental error on
$\sin (2\beta)$ is 
large, but is expected to be improved in the near future by the $B$ factories.
A determination of the angle $\beta$ from the measured $\sin (2\beta)$ still
has a four-fold ambiguity: $\beta$, $\pi/2-\beta$, $\pi+\beta$, $3\pi/2-\beta$
are all allowed solutions. Additional measurements are required to resolve the
discrete ambiguities \cite{kayser,charles,dighe,chiang,gross,quinn}. 
A major experimental effort will be made to determine the 
angles of the unitarity triangles \cite{stone}. It is also important to
precisely measure the magnitudes of the CKM matrix
elements, since measuring the phase angles alone cannot test
unitarity of the CKM matrix as discussed above.

\begin{table}[t]
\caption{Qualitative comparison of the sides of the unitarity triangles} 
\begin{tabular}{cccc} 
Relative length & Length    & Conventional triangle & New triangle \\ \hline
$1/1/1$ & $\lambda^3,\lambda^3,\lambda^3$ & $tu$, $bd$ & $ub$, $td$ \\ \hline
$1/1/\lambda^2$ & $1,1,\lambda^2$ &  & $tb$ \\
   & $\lambda^2,\lambda^2,\lambda^4$ & $ct$, $sb$ & $cb$, $ts$ \\ \hline
$1/1/\lambda^4$ & $1,1,\lambda^4$ &  & $ud$ \\ 
   & $\lambda,\lambda,\lambda^5$ & $uc$, $ds$ & $us$, $cd$ \\ \hline
$1/1/\lambda^6$ & $1,1,\lambda^6$ &  & $cs$ 
\end{tabular} \vspace{6pt}
\label{tab:shape}
\end{table}

We can analyze the shapes of the unitarity triangles by using the Wolfenstein
parametrization \cite{wolf} of the CKM matrix. The sides of the 
unitarity triangles are compared in Table \ref{tab:shape}. The unitarity
triangles can be divided into two categories: fat and flat. 
The $ub$ and $td$ triangles, like the conventional $tu$ and $bd$ triangles,
have a ``fat shape'' --- all three sides of each of them are of comparable 
lengths. The other (conventional and new) triangles are flat --- they almost
collapse to a line, with two sides having comparable lengths and the other
side being suppressed relative to them by $O(\lambda^2)$, $O(\lambda^4)$,
and $O(\lambda^6)$, respectively. Among the new unitarity triangles,
the $ub$ and $td$ triangles may therefore be particularly useful for testing 
the standard model. Since all three sides of each of them have comparable
lengths, of order $\lambda^3$, measurements of the sides with only moderate 
precision would be sufficient to test the unitarity relations.  
In contrast,
the other unitarity triangles require high precision measurements of the sides
to make a test. Neglecting $\chi$ and $\chi^\prime$, we can deduce from
Eqs.~(\ref{eq:new}) and (\ref{eq:5eqs}) the unitarity relations 
\begin{eqnarray}  
\cos \beta &\simeq& \cos (\pi-\omega_{ub}) = 
\frac{|V_{cd}V_{ts}|^2+|V_{cs}V_{td}|^2-
|V_{ub}|^2}{2|V_{cd}V_{ts}V_{cs}V_{td}|}\,\, , \label{eq:ub} \\
\cos \gamma &\simeq& \cos (\pi-\omega_{td}) = 
\frac{|V_{us}V_{cb}|^2+|V_{ub}V_{cs}|^2-
|V_{td}|^2}{2|V_{us}V_{cb}V_{ub}V_{cs}|}\,\, , 
\label{eq:td}
\end{eqnarray} 
for the $ub$ and $td$ triangles, respectively. 
Similarly, from Eqs.~(\ref{eq:5eqs}), (\ref{eq:new1}) and (\ref{eq:new2}) we
can write explicitly for the $ub$ triangle
\begin{eqnarray}
\cos \frac{\beta}{2} &\simeq& \sin \frac{\omega_{ub}}{2} = 
\sqrt{\frac{s_{ub}(s_{ub}-|V_{ub}|)}{|V_{cd}V_{ts}V_{cs}V_{td}|}}\,\,\, , 
\label{eq:ub1} \\
\sin \frac{\beta}{2} &\simeq& \cos \frac{\omega_{ub}}{2} = 
\sqrt{\frac{(s_{ub}-|V_{cd}V_{ts}|)(s_{ub}-|V_{cs}V_{td}|)}
{|V_{cd}V_{ts}V_{cs}V_{td}|}}\,\,\, ,
\label{eq:ub2}
\end{eqnarray}
where $s_{ub}=(|V_{cd}V_{ts}|+|V_{cs}V_{td}|+|V_{ub}|)/2$; for the $td$
triangle 
\begin{eqnarray}
\cos \frac{\gamma}{2} &\simeq& \sin \frac{\omega_{td}}{2} = 
\sqrt{\frac{s_{td}(s_{td}-|V_{td}|)}{|V_{us}V_{cb}V_{ub}V_{cs}|}}\,\,\, , 
\label{eq:td1} \\
\sin \frac{\gamma}{2} &\simeq& \cos \frac{\omega_{td}}{2} =
\sqrt{\frac{(s_{td}-|V_{us}V_{cb}|)(s_{td}-|V_{ub}V_{cs}|)}
{|V_{us}V_{cb}V_{ub}V_{cs}|}}\,\,\, ,
\label{eq:td2}
\end{eqnarray}
where $s_{td}=(|V_{us}V_{cb}|+|V_{ub}V_{cs}|+|V_{td}|)/2$. A comparison
of these determinations of $\beta$ and $\gamma$ with the measured ones would
offer an excellent test of the standard model.

It is useful to have as many different tests of the CKM mechanism for quark
mixing and CP violation
as possible, since that allows many potential new physics effects to be 
explored and the features of new physics to be identified. The nine new
unitarity triangles complement the conventional ones,
providing more ways of testing the CKM mechanism
and probing new physics. These can be done by testing each of the
individual unitarity relations represented as a unitarity triangle alone and
by comparing the areas of the unitarity triangles with one another which are 
all equal in the standard model. These consistency checks may give us some clue
to develop some principles which might enable us to understand quark mixing
and CP violation beyond the phenomenological level.

\acknowledgments
This work was supported by the Australian Research Council.

{\tighten

} 


\begin{references}
\bibitem{km} M. Kobayashi, T. Maskawa, Prog. Theor. Phys. 49 (1973) 652.

\bibitem{cabi} N. Cabibbo, Phys. Rev. Lett. 10 (1963) 531.

\bibitem{gim} S.L. Glashow, J. Iliopoulos, L. Maiani, 
Phys. Rev. D 2 (1970) 1285.

\bibitem{triangle} L. L. Chau, W. Y. Keung, Phys. Rev. Lett. 53 (1984) 1802;\\
M. Gronau, J. Schechter, Phys. Rev. Lett. 54 (1985) 385;\\
M. Gronau, R. Johnson, J. Schechter, Phys. Rev. D 32 (1985) 3062;\\
J.D. Bjorken, I. Dunietz, Phys. Rev. D 36 (1987) 2109;\\
C. Jarlskog, R. Stora, Phys. Lett. B 208 (1988) 268;\\
C. Jarlskog, in {\it CP Violation}, edited by C. Jarlskog (World Scientific,
Singapore, 1988), p.~3;\\
J. L. Rosner, A. I. Sanda, M. P. Schmidt, in {\it Proceedings of the Workshop 
on High Sensitivity Beauty Physics at Fermilab}, Batavia, Illinois, USA, 1987, 
edited by A. J. Slaughter, N. Lockyer, and M. Schmidt (Fermilab, Batavia,
1988), p.~165;\\ 
C. Hamzaoui, J.L. Rosner, A.I. Sanda, ibid., p.~215;\\
J.D. Bjorken, Phys. Rev. D 39 (1989) 1396.

\bibitem{jarlskog} C. Jarlskog, Phys. Rev. Lett. 55 (1985) 1039.

\bibitem{alek} R. Aleksan, B. Kayser, D. London, 
Phys. Rev. Lett. 73 (1994) 18. 

\bibitem{silva} J.P. Silva, L. Wolfenstein, Phys. Rev. D 55 (1997) 5331.

\bibitem{PDG} Particle Data Group, D.E. Groom, {\it et al.}, 
Eur. Phys. J. C 15 (2000) 1.

\bibitem{beta} OPAL Collaboration, K. Ackerstaff {\it et al.},
Eur. Phys. J. C 5 (1998) 379;\\
CDF Collaboration, T. Affolder {\it et al.}, Phys. Rev. D 61 (2000) 072005;\\
ALEPH Collaboration, R. Barate {\it et al.}, hep-ex/0009058;\\
BELLE Collaboration, H. Aihara, hep-ex/0010008;\\
BABAR Collaboration, D.G. Hitlin, hep-ex/0011024.

\bibitem{bisa} A.B. Carter, A.I. Sanda, Phys. Rev. D 23 (1981) 1567;\\
I.I. Bigi, A.I. Sanda, Nucl. Phys. B 193 (1981) 85.

\bibitem{kayser} Y. Azimov, Phys. Rev. D 42 (1990) 3705;\\
B. Kayser, L. Stodolsky, hep-ph/9610522;\\
B. Kayser, in {\it Proceedings of the 32nd Rencontres de Moriond:
Electroweak Interactions and Unified Theories}, Les Arcs, France, 1997,
edited by J. Tran Thanh Van (Editions Frontieres, 1997), p.~389, 
hep-ph/9709382.

\bibitem{charles} J. Charles, A. Le Yaouanc, L. Oliver, O. Pene, J.-C. Raynal,
Phys. Lett. B 425 (1998) 375; 433 (1998) 441(E).

\bibitem{dighe} A. Dighe, I. Dunietz, R. Fleischer, 
Eur. Phys. J. C 6 (1999) 647.

\bibitem{chiang} C.-W. Chiang, L. Wolfenstein, Phys. Rev. D 61 (2000) 074031.

\bibitem{gross} Y. Grossman, D. Pirjol, J. High Energy Phys. 006 (2000) 029.

\bibitem{quinn} H.R. Quinn, T. Schietinger, J.P. Silva, A.E. Snyder,
hep-ph/0008021.

\bibitem{stone} S. Stone, hep-ph/9910417.

\bibitem{wolf} L. Wolfenstein, Phys. Rev. Lett. 51 (1983) 1945.
\end{references}
\end{document}